# Dilated space-and-wavelength selective crosspoint optical switch

ZIYAO ZHANG,[1] MINJIA CHEN,[1] RUI MA,[1] BOHAO SUN, [1] ADRIAN WONFOR,[1] RICHARD PENTY, [1] QIXIANG CHENG[1,2]*

[1]*Department of Engineering, Centre for Photonic Systems, Electrical Engineering Division, University of Cambridge, Cambridge CB3 0FA, UK*
[2]*GlitterinTech Limited, Xuzhou, 221000, China*
*\*qc223@cam.ac.uk*

**Photonic integrated switches that are both space and wavelength selective are a highly promising technology for data-intensive applications as they benefit from multi-dimensional manipulation of optical signals. However, scaling these switches normally poses stringent challenges such as increased fabrication complexity and control difficulties, due to the growing number of switching elements. In this work, we propose a new type of dilated crosspoint topology, which efficiently handles both space and wavelength selective switching, while reducing the required switching element count by an order of magnitude compared to reported designs. To the best of our knowledge, our design requires the fewest switching elements for an equivalent routing paths number and it fully cancels the first-order in-band crosstalk. We demonstrate such an ultra-compact space-and-wavelength-selective switch (SWSS) at a scale of 4×4×4λ on the silicon-on-insulator (SOI) platform. Experimental results reveal that the switch achieves an insertion loss ranging from 2.3 dB to 8.6 dB and crosstalk levels in between -35.3 dB and -59.7 dB. The add-drop microring-resonators (MRRs) are equipped with micro-heaters, exhibiting a rise and fall time of 46 µs and 0.33 µs, respectively. These performance characteristics highlight the switch's ultra-low element count and crosstalk with low insertion loss, making it a promising candidate for advanced data center applications.**

## 1. Introduction

Data traffic demand in data centers is experiencing exponential growth with the advent of Internet live streaming, artificial intelligence (AI) and other data-intensive technologies [1]. Optical switches offer a promising solution to meet the network capacity requirements due to their high bandwidth, low power consumption, and low crosstalk [2, 3]. Optical switches have been investigated using various techniques, such as micro-electromechanical systems (MEMS) [4-6], liquid crystal [7, 8], and integrated photonics [9, 10]. Among these, silicon on insulator (SOI) integrated optical switches demonstrate significant potential due to their small footprint, and fast reconfiguration speed, and CMOS compatibility.

So far, most photonic integrated switches are designed to be either space selective, directing optical signals through distinct spatial paths, or wavelength selective, routing signals by assigning specific wavelengths to predefined paths. For example, in 2019, Cheng et al. demonstrated a microring-resonator (MRR) based switch-and-select space switch with ultra-low crosstalk [11]. Later in 2023, Gao et al. scaled such a design to a 32 × 32 port count on a multi-layer $Si_3N_4$-on-SOI platform [12]. On the other hand, as a key element of reconfigurable optical add/drop multiplexers (ROADMs), wavelength-selective switches (WSSs) have also seen significant advancements. In 2022, Vagionas et al. reported a 1×4×4λ WSS based lossless ROADM [13]. In 2023, Zhang et al. demonstrated a 1×4×4λ WSS [14], utilizes the adiabatic elliptical MRRs, with low on-chip loss and low crosstalk.

However, by leveraging both spatial and wavelength selectivity, the capacity of a single optical switch fabric can be significantly enhanced. This advancement would not only improve the efficiency of bandwidth allocation but also provide exceptional flexibility in network configuration, enabling seamless management of massive data flows that overwhelm existing systems [2]. Several studies have proposed designs that can simultaneously achieve both spatial and spectral selectivity [15]. Some designs rely on arrayed waveguide gratings (AWGs) that could pose challenges in fabrication [16-18]. On the other hand, designs with MRRs offer a smaller footprint and are wavelength-sensitive for spatial add-drop, being ideal for space-and-wavelength selective switches (SWSSs). In 2019, Cheng et al. proposed a scalable architecture featuring MRR-based wavelength selectors and comb aggregators [19], and the simulation results indicate this architecture supports up to 192 connectivity (16 ports × 12λ). In the same year, Khope et al. demonstrated an 8×4×2λ crossbar SWSS [20] with minimum and maximum on-chip path losses of 6 dB and 14 dB, respectively, and with out-of-band rejection of -21 dB and -32 dB at 200 and 400 GHz channel spacing, respectively. More recently, in 2024, Luo et al. explored the potential of all-pass MRRs as phase shifters for SWSSs

[21]. In this design, Mach-Zehnder Interferometers (MZIs) with multiple pairs of MRR phase shifters act as the switching elements, and an 8×8×8λ SWSS is theoretically verified.

These reported approaches advance this field; however, all face a fundamental scaling issue: the number of switching elements required grows cubically as the connectivity scales up. This limits the scalability and practicality of current designs, especially for large-scale data-centre interconnects, where high port counts and large numbers of wavelengths are essential. The challenge lies in developing a solution that can support simultaneous spatial and wavelength selectivity without the prohibitive cost and complexity that accompany scaling up the number of switching elements. In this paper, we present a new type of dilated crosspoint topology and demonstrate an ultra-compact SWSS at the scale of 4×4×4λ on the SOI platform. This innovative design achieves both space and wavelength selective switching, with substantially reduced switching element count, transitioning from cubic growth to quadratic growth as the connectivity scales up. In addition, this topology is immune to first order in-band crosstalk. Our 4×4×4λ SWSS requires only 32 MRRs to control 64 routing paths with a small footprint of only 2.25 mm². This integrated dilated crosspoint optical switch, featuring ultra-compact footprint, low insertion loss and low crosstalk, shows significant potential for optical interconnects in data centers.

## 2. Design and Fabrication
### a. Topology

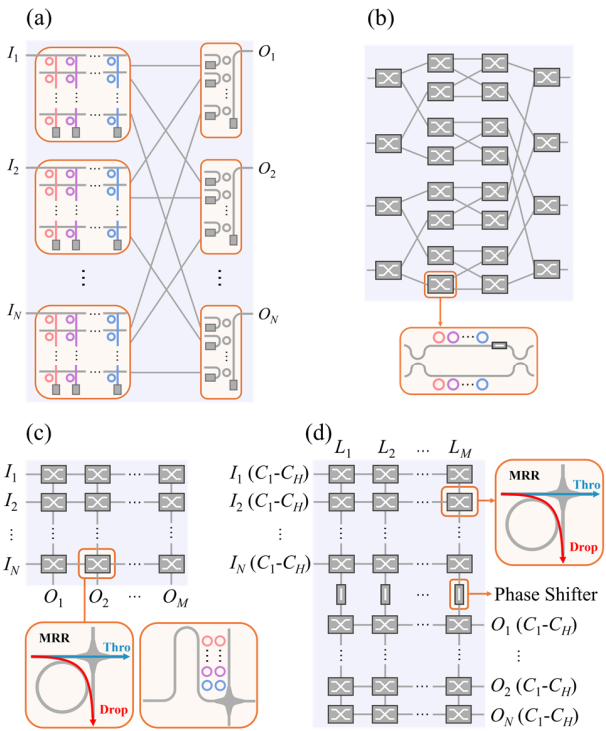

**Fig. 1.** Diagrams of the SWSS topologies. (a) Switch and selective SWSS featuring MRR-based wavelength selectors. (b) Dilated Banyan topology with MRRs assisted MZI as switching elements. (c) Traditional crossbar topology and (d) dilated crosspoint topology proposed in this work.

Figure 1 illustrates state-of-the-art SWSS topologies that have been reported to date. Figure 1(a) is the switch and selective SWSS featuring MRR-based wavelength selectors [19]. Figure 1(b) presents a dilated Banyan topology with MRRs-assisted MZIs as switching elements [21]. Figure 1(c) depicts the traditional crossbar topology, a non-blocking spatially selective switching configuration that has been extensively developed [22-24]. Ref. [20] proposed a crossbar SWSS using a switching element consisting of a series of second-order MRRs to replace the single MRR, achieving wavelength selectivity and crosstalk suppression.

Figure 1(d) shows the schematic of the proposed dilated crosspoint topology that utilizes wavelength-division multiplexing (WDM) technology to achieve both wavelength and spatially selective switching. The topology consists of two stages of identical cascaded traditional crossbar structures, with phase shifters connecting the outputs of the first stage to the inputs of the second stage. Each signal passes through two on-state MRRs and a phase shifter before reaching the target output port. The topology comprises $M$ columns, $L_1$-$L_M$, and $M \geq H$ is necessary to route signals in $H$ wavelength channels ($C_1$-$C_H$) from $N$ input ports to $N$ output ports simultaneously. Crosstalk from the same input port to the same target output port via different paths may have identical optical path lengths under a few very specific scenarios, which amplifies interference-related crosstalk. For example, as illustrated in Fig. 2, signals represented by purple solid line are routed from $I_3$ to $O_3$. If the leakage signal from other off-state rings, represented by dashed lines, shares the same optical path length, constructive interference occurs and hence enhances crosstalk. To prevent this situation, the four phase shifters are kept at phases of 0, $\pi/2$, $\pi$, and $3\pi/2$, respectively, avoiding unnecessary interference.

We define a permutation as a specific connection from an input port to a designated output port within a channel, irrespective of the internal routing paths. To explore the blocking theory of the dilated crosspoint topology, we calculate the global switching states, which are defined as the total number of switch configurations that result in valid routing permutations. Wide-sense and strictly non-blocking networks both enable paths to be established between idle inputs and outputs without disrupting existing connections. The difference lies in that the former adheres to specific routing rules, while the latter imposes no such constraints. These networks are commonly preferred for their simplified switching control system. Since the traditional crossbar topology is wide-sense non-blocking, we thus investigate the wide-sense non-blocking condition for this SWSS.

For simplicity, we assume $M=N=H$, which transforms each stage of the crossbar into a square matrix. In this topology, the routing principle requires that MRRs in the same rows and columns of each stage of the crossbar must not operate on the same wavelength channel to avoid blocking. To simplify the analysis, each stage of the crossbar is represented as a square matrix where each element corresponds to a wavelength channel number of an MRR's resonant wavelength. This matrix, where each number appears exactly once in each row and column, is known as a Latin Square [25]. While no exact formula exists for determining the number of $N \times N$ Latin Square $L_N$, the exact counts for $N \leq 11$ have been computed and reported, and approximate counts for $12 \leq N \leq 15$ have also been provided [25].

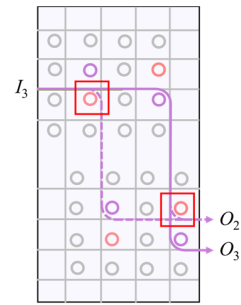

**Fig. 2.** An example of crosstalk increased by interference.

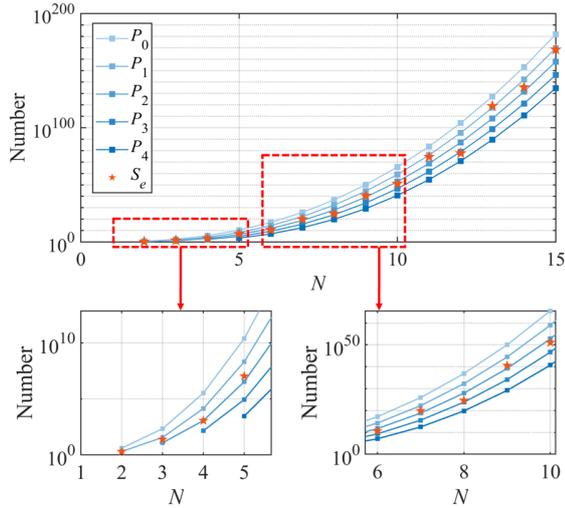

**Fig. 3.** Permutation number and effective switching states as functions of port number $N$. Different permutation numbers with different wavelength channels are compared. $S_e > P_M$ indicates that the switch meets the wide-sense non-blocking condition.

**Table 1. Wide-sense non-blocking condition for dilated crosspoint topology with different ports number**

| $N$ | Maximum Number of Channels |
|---|---|
| 2, 13 | $N-1$ |
| 3, 4, 5, 7, 9, 11, 14, 15 | $N-2$ |
| 6, 8, 10 | $N-3$ |
| 12 | $N-4$ |

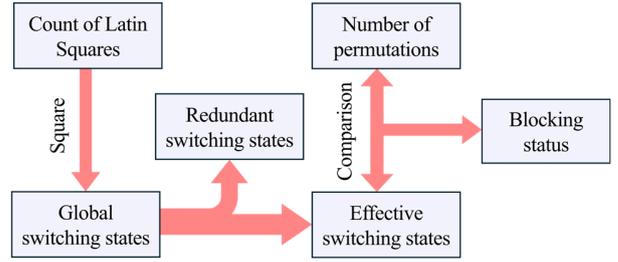

**Fig. 4.** Flow chart of the blocking status calculation process.

Among these global switching states, some configurations lead to identical permutations, termed redundant switching states. Consequently, even if the number of global switching states meets or exceeds the number of permutations, the switch may still exhibit blocking behavior. Thus, we exclude all redundant switching states, making it straightforward to determine the blocking status. First, we calculate the number of permutations among all the wavelength channels, under the assumption that

$$P(N) = (N!)^H = (N!)^N. \tag{1}$$

To calculate the global switching states, we obtain the count of Latin Squares $L_N$ from a look-up table [25]. Since each stage of the crossbar topology is represented by a Latin Square, the number of global switching states is

$$S(N) = L_N^2. \tag{2}$$

After excluding all redundant switching states, the remaining states are referred to as effective switching states, meaning each effective switching state corresponds to a unique permutation. The detailed calculation process for determining the effective switching states is provided in the Appendix A. The number of redundant switching states can be expressed as:

$$S_r(N) = \prod_{i \in D1}\left( \left(C_N^i \cdot i!\right) \cdot \prod_{j \in D2}\left(C_{\frac{N}{i}}^j \cdot j!\right) \right) \cdot \prod_{l \in H}\left(C_N^l \cdot l!\right). \tag{3}$$

In this equation, the set $D1$ consists of all divisors of $N$, and the set $D2$ consists of divisors of $N/i$, and the set $H$ consists of the positive integers smaller than $N/2$. All these sets exclude 1, and $H$ excludes the same elements in $D1$.

Effective switching states can be computed by $S$ and $S_r$:

$$S_e(N) = \frac{(S(N))^2}{S_r(N)} \tag{4}$$

**Table 2. The figures of merit for various reported SWSS topologies**

| Reference | Topology | Elements Number | Global Switching States | Order of Crosstalk | Number of Stages |
|---|---|---|---|---|---|
| Ref. [21] (NB) | Dilated Banyan, B&S | $(4N^2-4N) \cdot H$, $2HN^2+2\log_2 N$ | $(N!)^H$ | Second | $2\log_2 N$, $2\log_2 N+1$ |
| Ref. [19] (NB) | S&S | $(H+1)N^2+N^2$ | $(N!)^H$ | Second | $3N-1$ |
| Ref. [27] (NB) | PILOSS | $H \cdot (N^2+2N)$ | $(N!)^H$ | First | $N^2$ |
| Ref. [20,26] (NB) | Crossbar | $HN^2$ | $(N!)^H$ | First | $2N-1$ |
| Ref. [27] (FCB) | PILOSS | $H \cdot (N^2/4+N)$ | $((N/2)!)^H$ | First | $2N^2$ |
| This work (NB) | Dilated crosspoint | $2MN$ | Look-up table | Second | $4N-2$ |
| This work (FCB) | Dilated crosspoint | $2HN$ | Look-up table | Second | $4N-2$ |

NB: Non-blocking; FCB: Fully-connected blocking.
B&S: Broadcast-and-Select, S&S: Switch-and-Select.
$H$: Channel number; $N$: Port number.

Based on the calculated result where $P(N) > S_e(N)$, the switch is determined to be blocking, which means different wavelength allocations would conflict with spatial channels. To ensure non-blocking operation, the number of wavelength channels $H$ must be reduced to smaller than $N$. Consequently, the number of permutations is reduced to:

$$P_{N-H}(N, H) = (N!)^H. \quad (5)$$

Figure 3. illustrates the computed $P_{N-H}$ and $S_e$ as functions of $N$, when $N-H$ equals 0-4 respectively. $S_e > P_{N-H}$ indicates that the switch meets the wide-sense non-blocking condition, and all wavelength channels can be arbitrarily allocated without conflicting with spatial switching. According to Fig. 3, in Table 1, we list the maximum channel count for this SWSS at various values of $N$ to achieve a wide-sense non-blocking operation. The flowchart of the blocking status calculation process is shown in Fig. 4. Firstly, the total number of global switching states is calculated by squaring the count of Latin squares. Next, the effective switching states are obtained by excluding redundant states. Finally, by comparing the number of permutations with the effective switching states, we can determine whether the SWSS is blocking or wide-sense non-blocking. For more information about the blocking theory, please see the Appendix A.

Table 2. presents the figures of merit for various reported SWSS topologies [19-21, 26, 27]. To calculate the global switching states for the dilated crosspoint topology, which is also the count of Latin Square, a look-up table is necessary that is illustrated in the Ref. [25]. Compared to other SWSS topologies, the dilated crosspoint SWSS requires the fewest switching elements, on the order of $HN$, while others require elements on the order of $HN^2$. Therefore, this topology can significantly reduce the number of switching elements, especially as the SWSS scales up. Figure 5 illustrates the number of switching elements as functions of the port number $N$ (for $N>4$) for different reported SWSS topologies. For clarity, we have re-scaled the reported SWSS to the same port count and channel number as the dilated crosspoint topology. In the figure, triangle dots without lines represent strictly non-blocking SWSS, square dots without lines represent wide-sense non-blocking SWSS, and dots with dashed lines represent blocking SWSS. Although the blocking and wide-sense non-blocking dilated crosspoint SWSS are the same topology structure, they differ in the number of wavelength channels depicted in Fig. 3. This means that non-blocking SWSSs should only be compared with other non-blocking SWSSs, and similarly, blocking SWSSs should be only compared with other blocking SWSSs. Notably, for both kinds of SWSSs, this dilated crosspoint SWSS requires the fewest switching elements to achieve the same routing paths.

While the dilated crosspoint SWSS shows significant advantages in terms of reducing switching elements, it is also essential to consider signal integrity for practical applications. This proposed dilated crosspoint topology can fully cancel the first order in-band crosstalk. The input power is denoted as $\rho_0$, and the output power for the routing path from $I_i$ to $O_j$ is denoted as $\rho_{ij}$. The on-chip insertion loss of the path $I_i$-$O_j$ is derived as follows:

$$\varepsilon_{ij} = -10 \cdot \log_{10}(\rho_{ij}/\rho_0). \quad (6)$$

The detected leakage power to a non-targeting output-port $k$ ($k \neq j$) is represented as $\tau_{ijk}$, and the crosstalk can be calculated:

$$\xi_{ijk} = -10 \cdot \log_{10}(\tau_{ijk}/\rho_0). \quad (7)$$

Consequently, the crosstalk ratio $\chi_{ijk}$ is derived as

$$\chi_{ijk} = \xi_{ijk} - \varepsilon_{ij}. \quad (8)$$

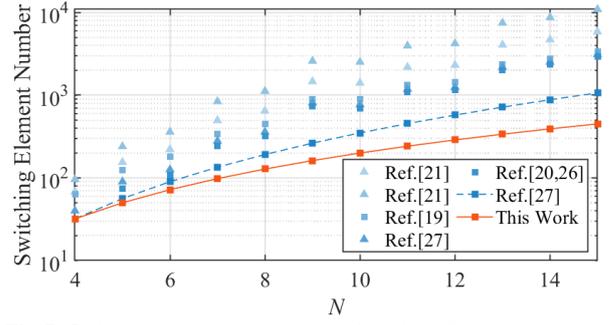

**Fig. 5.** Switching element number as a function of the port number $N$ among different reported SWSSs.

This formula quantifies the crosstalk by comparing the detected leakage power with the output power of the intended routing path.

**b. MRR design space exploration**

To optimize the switch performance, a design space exploration of the MRR is conducted using the theory and methodologies outlined in reference [28] and [29]. This approach allows for the selection of optimal MRR parameters, such as radius and gap, which not only achieve the desired performance but also ensure robustness against potential deviations caused by fabrication inaccuracies. The numerical simulations utilize bending loss values obtained from the experimental results presented in Ref. [29]. Figures 6 (a)-(c) depict the contours of the 3 dB bandwidth, attenuation at resonance, and extinction ratio (ER) at the drop port. In these contours, the x-axis shows the MRR radius and the y-axis shows the power coupling efficiency, which is defined as $\kappa^2$. According to the criteria outlined in reference [29], we further refine the MRR parameters to balance loss, crosstalk, and passband requirements in the context of our switch architecture. Specifically, the drop insertion loss is constrained to be lower than 1 dB to ensure efficient signal routing, while the extinction ratio for the MRR is maintained larger than 30 dB to provide at least 20 dB crosstalk suppression in a WDM-based link. Moreover, the passband is expanded to over 20 GHz to accommodate high-speed data transmission, ensuring compatibility with signalling rates greater than 20 Gbps per channel. These optimizations are achieved by exploring the design space, considering variations in MRR radius and coupling gap to ensure a balance between insertion loss, crosstalk suppression, and 3 dB bandwidth. Figure 6 (d) provides a comprehensive view of the design space, taking into account these three performance metrics, with the white region indicating where the MRRs meet all the specified criteria. The optimized MRRs are selected from the center of this white region, corresponding to a radius of 7 μm and a $\kappa^2$ of 0.04. Additionally, FDTD simulations are conducted to examine the power coupling efficiency as a function of the gap between the MRR and the bus waveguides for an MRR radius of 7 μm. Figure 6 (e) shows that a 0.17 μm gap corresponds to a power coupling efficiency of 0.04. Therefore, the optimized MRR has a radius of 7 μm and a gap of 0.17 μm.

**c. SWSS design**

A 4×4×4λ SWSS based on the dilated crosspoint topology is demonstrated in this work, and its schematic is illustrated in Fig. 7 (a). The switch is designed on an SOI platform, comprising two 4×4 crossbar structures. As mentioned in Section 2, four phase shifters interconnect the first-stage crossbar and the second-stage crossbar, serving to avoid crosstalk enhancement by interference, through tuning to 0, π/2, π and 3π/2 respectively. Add-drop MRRs function as switching elements in this SWSS, and there are 32 MRRs in total, for controlling 64 routing paths in 4 different wavelength channels. The diagram of a single switching element is shown in Fig. 7 (b). The MRR is thermally-tuned via a micro-heater and the optical signal propagates directly to the through port when the MRR is in the off-state at a specific wavelength. Conversely, when the MRR is tuned to the on-state the signal is directed to the drop-port, which is located on the vertically crossing waveguide.

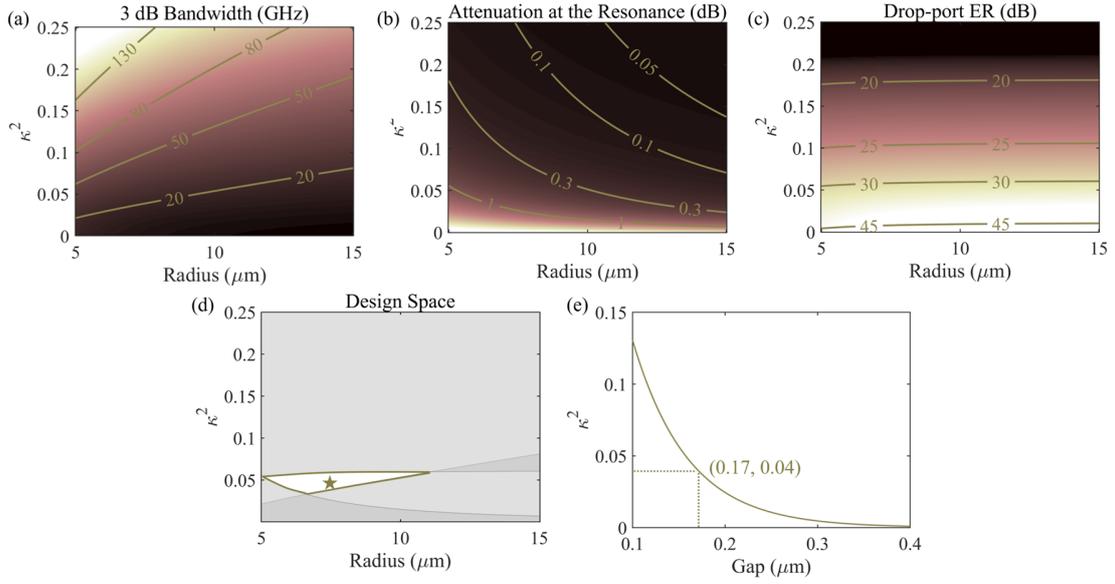

**Fig. 6.** Design space exploration of the MRR. (a) Contour of the 3 dB bandwidth versus MRR radius and $\kappa^2$. (b) Contour of the attenuation at the resonance. (c) Contour of the drop-port extinction ratio. (d) Overall design of the MRR. (e) Power coupling efficiency $\kappa^2$ as a function of the MRR gap.

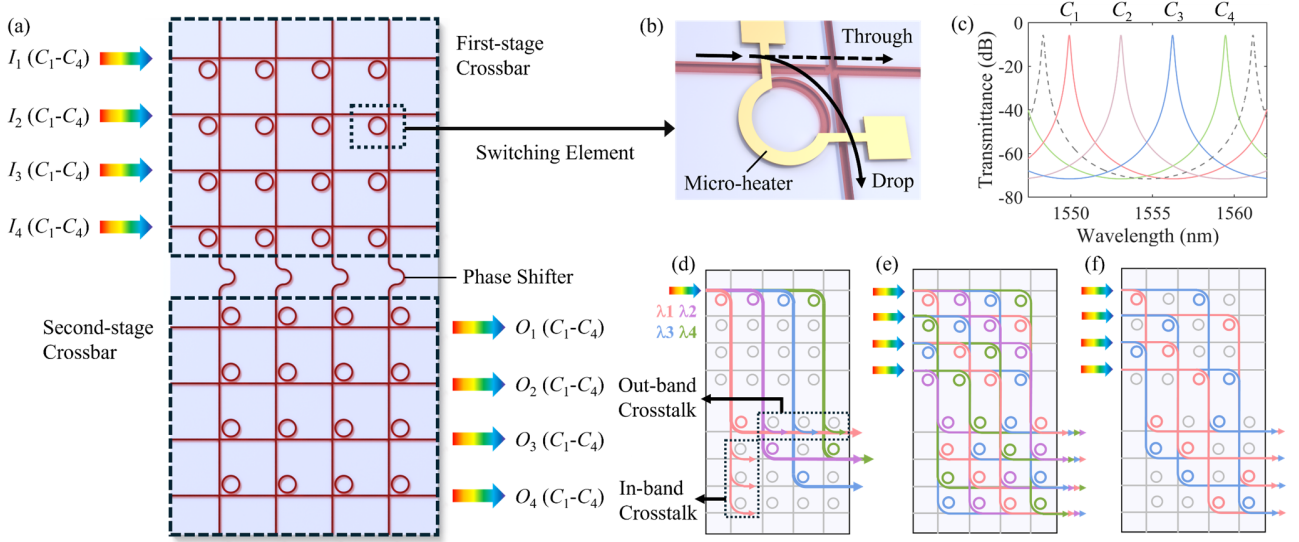

**Fig. 7.** (a) Schematic of the 4×4×4λ dilated crossbar optical switch. (b) 3D model of the switching element. (c) Simulated transmission spectrum of the four wavelength channels. (d) An example showing signals in four wavelength channels are routed from $I_1$ to $O_1$, $O_2$, $O_3$, and $O_2$, respectively. The second order in-band crosstalk and the first order out-band crosstalk in channel 1 is illustrated. (e) Diagram of the fully-connected blocking SWSS operating under a fully loaded condition. (f) Diagram of the wide-sense non-blocking SWSS operating under a fully loaded condition.

This optical switch features four wavelength channels: 1549.9 nm, 1553.1 nm, 1556.3 nm and 1559.5, as shown in Fig. 7 (c). In this figure, the grey dashed line represents the spectrum of the path comprised of the de-energized MRRs, and these wavelength channels are uniformly arranged within the MRR's free spectral range (FSR). As illustrated in Section 2, this 4×4×4λ SWSS is a fully-connected blocking switch, and can also work as a 4×4×2λ SWSS wide-sense non-blocking switch. During operation, the MRRs can be dynamically tuned to resonance in the corresponding wavelength channels, allowing the signals in these four channels to be routed from specific input ports to any selected output ports.

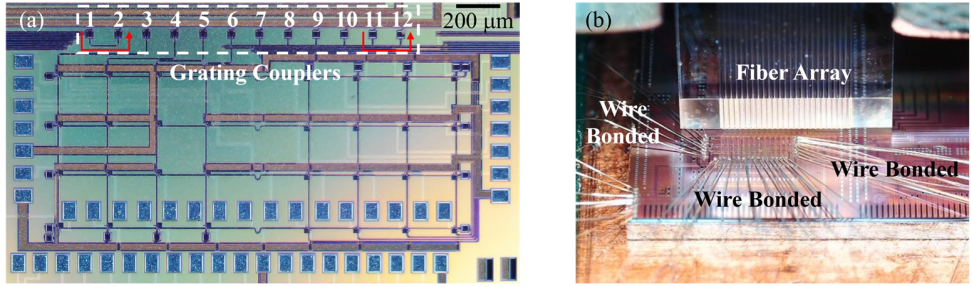

**Fig. 8.** (a) Microscopic photograph of the dilated crossbar optical switch. (b) Photograph of the photonic chip with wire bonding and optical coupling with ultra-high numerical aperture (UHNA) fiber array.

Figure 7 (d) illustrates an example where signals in four wavelength channels that are marked as different colours are routed from $I_1$ to $O_1$, $O_2$, $O_3$, and $O_2$ simultaneously. Signals in each wavelength channel are routed to the target output-port by two on-state MRRs. The primary source of crosstalk in this switch is second-order in-band crosstalk and first-order out-band crosstalk, and the crosstalk in channel 1 is indicated by the thin arrows in Fig. 7 (d). Figure 7 (e) presents an example of the fully-connected blocking switch operating under a fully loaded condition, in which all the four input-ports, output-ports and wavelength channels are occupied. Correspondingly, Figure 7 (f) is an instance of the wide-sense non-blocking switch operating under a fully loaded condition.

### d. Fabrication

This integrated optical switch is fabricated by the Interuniversity Microelectronics Centre (IMEC) on an SOI platform, as depicted in the microscope image shown in Fig. 8 (a). This compact SWSS has a footprint of only 2.25 mm². Each input-port and output-port correspond to grating couplers aligned in a line with a pitch of 127 µm. In addition to the four input ports and four output ports (ports 3-10), there are two additional pairs of grating couplers (ports 1, 2, 11, 12) at each end used for optical alignment, as indicated by the red arrows in Fig. 8 (a). The SWSS includes 36 control electrodes along with three common grounds. As illustrated in Fig. 8 (b), the chip is wedge-bonded for control purposes.

## 3. Result

### a. Experiment set up

Figure 9 illustrates the diagram of the SWSS testing setup. A C-band tunable laser is used to emulate four input wavelength channels by sequentially adjusting its output wavelength. An ultra-high numerical aperture (UHNA) fiber array is employed for light coupling. A polarization controller is connected between the tunable laser and the UHNA fiber array to align the polarization state of optical input to the chip, thereby reducing the mode mismatch between the UHNA fiber array and the grating couplers. For the chip tuning, a PCB board is connected to the chip by wedge bonding. The output power is measured using an optical power meter. The tunable laser, the optical power meter, and the PCB board are all controlled by a computer. In Fig. 9, the optical connections are represented by the solid lines, and the electrical connections are represented by the dashed lines.

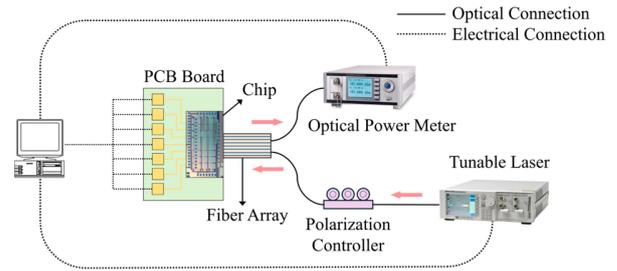

**Fig. 9.** Diagram of the SWSS testing setup.

### b. On-chip Loss and Crosstalk

Initial measurements are of the crosstalk of the SWSS. As an example, the routing path from $I_4$ to $O_1$ is evaluated with signals in all four wavelength channels. The tunable laser wavelength is swept across the wavelength region of the four wavelength channels, and the output power is recorded by an optical power meter. Figure 10 (a) shows the spectrum of this path when all four wavelength channels are tuned to the on-state. Four wavelength channels are distinguished by different background colours. These channels have a spacing of 3.2 nm, and an extinction ratio (ER) of 38.8 dB.

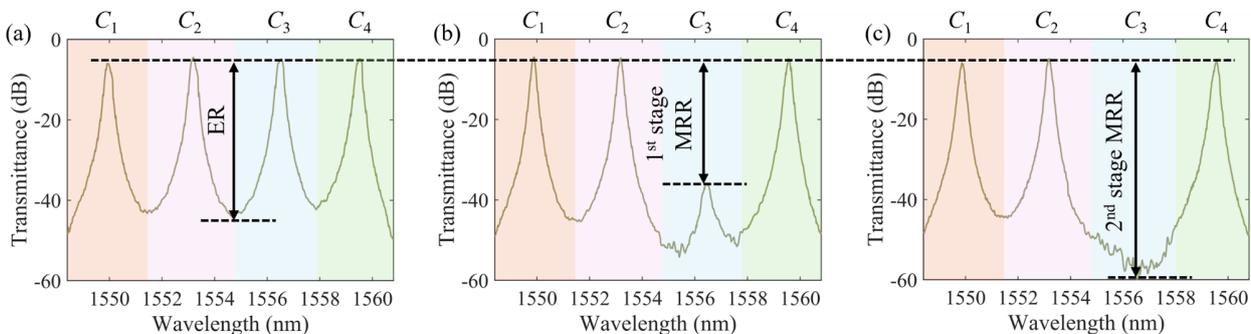

**Fig. 10.** Measured spectra of routing path from $I_4$ to $O_1$. (a) Spectrum when all paths are on-state. (b) Spectrum when wavelength channel 3 is at off-state with first-order crosstalk. (c) Spectrum when wavelength channel 3 is off-state with second-order crosstalk.

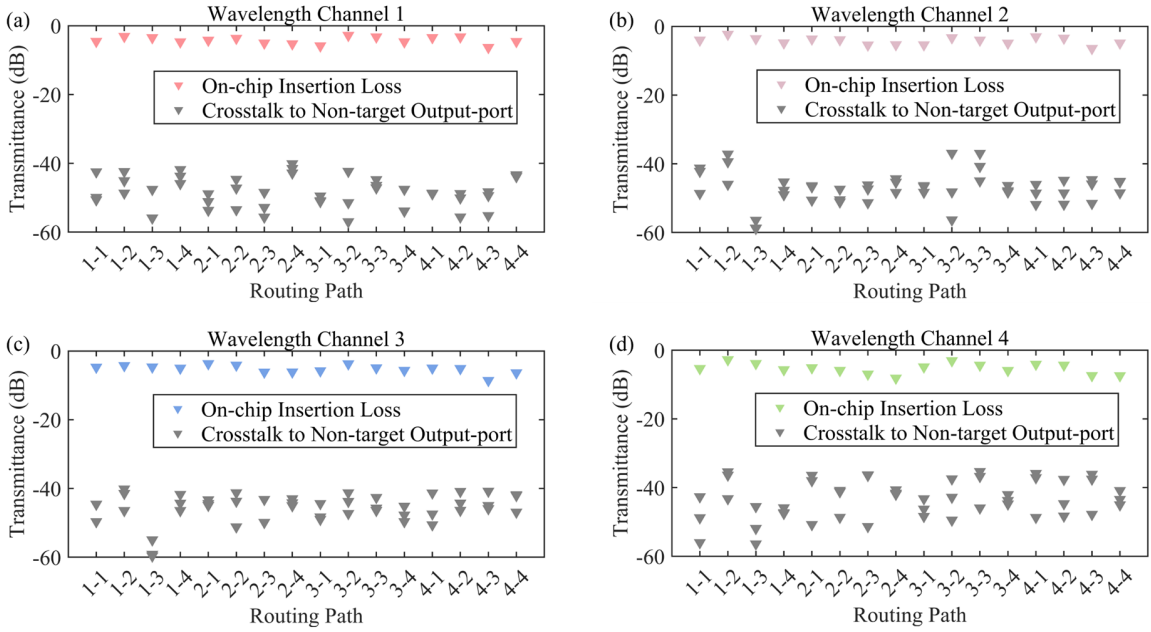

**Fig. 11.** Measured optical power map of the optical switch for four wavelength channels.

To analyze the first-order leakage, one of the MRRs on the path is tuned to the off-state, allowing only crosstalk to leak to the non-target output port ($O_1$ in this instance) from the drop port of the off-state MRR. Figure 10 (b) shows the first-order crosstalk ratio for wavelength $C_3$ when the corresponding MRR is tuned to the off-state. In this case, the off-state MRR is tuned to resonance in wavelength $C_1$, corresponding to the off-state of $C_3$. Additionally, to assess the 2nd order leakage, the two MRRs on the path of the $C_3$ are tuned to the off-state, and the spectrum is shown in Fig. 10 (c).

For this instance, one of the MRR is tuned to resonance in $C_1$ and the other in $C_2$. The testing results indicate that for this routing path, the first-order crosstalk ratio is -31.0 dB and the second-order crosstalk ratio is -55.2 dB.

The optical power maps are further tested, indicating the on-chip losses and crosstalk of all 64 routing paths of this 4×4×4λ optical switch, and the corresponding results are shown in Fig. 11 (a-d). In these figures, the colored dots represent the insertion loss, and the gray dots represent the crosstalk leakage. The optical power maps for different wavelength channels are displayed separately in four sub figures. To obtain the crosstalk leakage to non-target output ports, the MRRs responsible for the crosstalk leakage as illustrated in Fig. 7 (d), are tuned to the off-state. Although the crosstalk enhancement induced by interference as illustrated in Section 2 and Fig. 2 does not exist in our testing, we still activate the phase shifters as a rigorous test procedure. The results show that the insertion loss varies from 2.3 dB to 8.6 dB, while the crosstalk ranges between -35.3 dB and -59.7 dB. Path-dependent loss is not an advantageous feature of the Crossbar topology, and the MRR switch elements could make it worse as there is a clear discrepancy in its drop and through loss, such as the results reported in Ref. [11, 20, 37]. The variation in insertion loss can be mainly attributed to the difference in waveguide crossings (evaluated at 0.2 dB per crossing), waveguide lengths and off-state MRRs together with wavelength-dependent off-state losses. In practice, the de-energized MRRs can be de-tuned as illustrated in Fig. 7 (b) to mitigate the loss variations. Furthermore, variable optical attenuators could be applied to further equalize the path-dependent loss if it is a primary design consideration [30].

Deriving from the power map, the crosstalk ratios for all routing paths, with leakage to all non-target output ports are depicted in Fig. 12 as heat maps. Similar to the optical power map, the four wavelength channels are represented by distinct colours. The crosstalk ratios for these paths range from -28.7 dB to -55.6 dB.

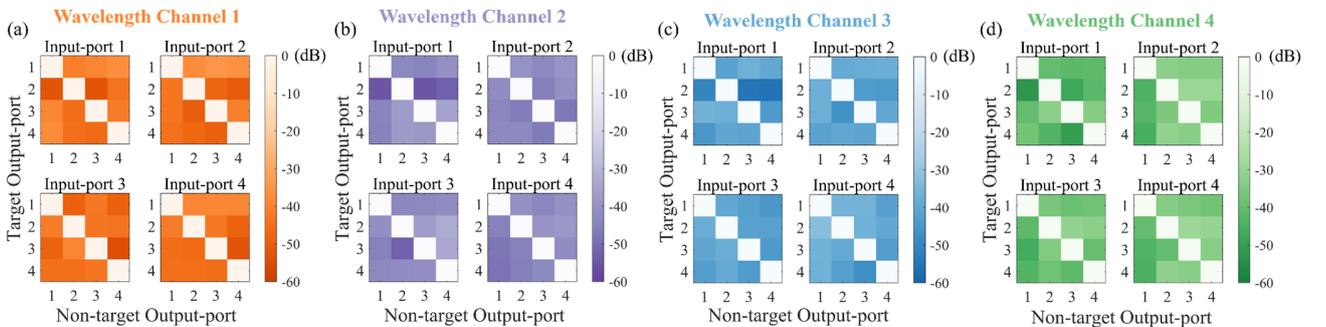

**Fig. 12.** The crosstalk ratios for all routing paths with leakage to all non-target output ports. The four wavelength channels are represented by distinct colours.

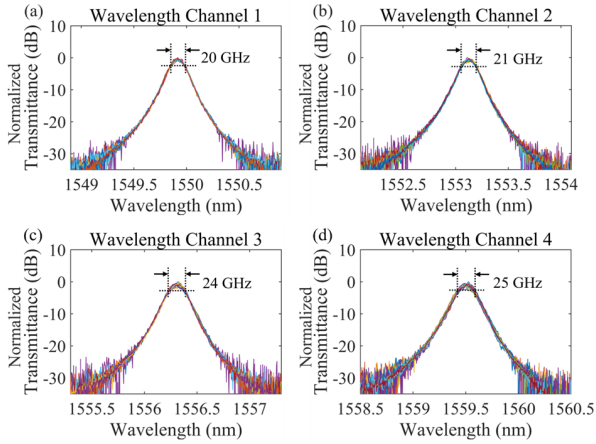

**Fig. 13.** The superimposed spectrum of all routing permutations across four channels. Each sub-figure illustrates all 16 possible permutations within each wavelength channel.

### c. 3-dB bandwidth

Additionally, the 3-dB bandwidths of the SWSS are characterized. A C-band superluminescent diode (SLD) is used as the broadband optical source, and a spectrometer with a resolution of 0.04 nm is utilized to record the spectrums. The spectra for all paths across the four channels are normalized to the same benchmark, using the spectrum of path $I_1$-$O_1$ as the reference. The superimposed spectra are shown in Fig. 13. The measured 3 dB bandwidths for the four channels are around 20 GHz, 21 GHz, 24 GHz, and 25 GHz, respectively. The 3 dB bandwidth in different channels differs, as explained in Ref. [28]. The formula for Full Width at Half Maximum (FWHM) is:

$$\text{FWHM} = \frac{(1-ra)\lambda_{res}^2}{\pi n_g L \sqrt{ra}} \quad (9)$$

From this equation, it can be seen that a larger resonance wavelength leads to a larger FWHM, which corresponds directly to the 3 dB bandwidth.

### d. Switching time

The optical reconfiguration time of the thermo-optic switch is evaluated by measuring its time-domain response. A 2 kHz electrical square-wave signal with a 50% duty cycle is applied to the path $I_1$-$O_4$, and the temporal trace of the optical power is recorded by an oscilloscope, shown as Fig. 14. The rise and fall times of the switch are tested to be 47.58 μs and 0.33 μs respectively, measured between the 10% and 90% power points of the response. The imbalance in switching rise and fall times, as also observed in many other publications such as Ref. [11], is primarily attributed to the chip's efficient thermal conductivity [31] and the specific design of the heater. The rapid heat dissipation leads to a much shorter fall time but the propagation speed of heat severely limits the rise time. One potential solution is to introduce thermal isolation trenches to ensure more controllable and balanced heating and cooling dynamics, also help reduce thermal crosstalk between neighbouring components [32]. This however requires additional fabrication processes.

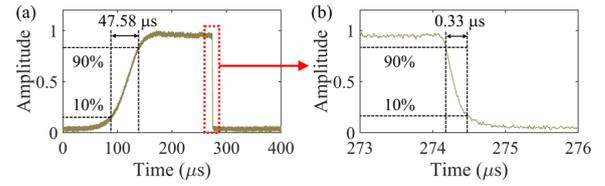

**Fig. 14.** Reconfiguration time of the SWSS.

## 4. Discussion

Table 3. illustrates the performance comparison between this SWSS and other reported integrated switches [11, 14, 18, 20, 33]. In addition to requiring fewer switching elements to achieve the same scale, this SWSS also shows low on-chip loss and low crosstalk, especially compared with other WSSs and SWSSs. Future work is expected to bring further improvements in switching performance. First, replacing the TO switching elements with electro-optic (EO) elements [34, 35] could enhance switching times from the microsecond to the nanosecond range, significantly increasing data transmission bandwidth. Second, utilizing higher-order MRRs can achieve a flat-top transmission spectrum, and substantially reduce crosstalk [20, 36]. At the same time, by leveraging higher-order designs, it becomes flexible to expand the 3 dB bandwidth, thereby hosting high-speed data transmission systems. In 2014, DasMahapatra et al. introduced a fifth-order MRR-based optical switch with a 3 dB bandwidth of up to 100 GHz [37]. Later, in 2020, Huang et al. reported a switch utilizing a dual-MRR configuration, achieving a 3 dB bandwidth of 165 GHz [38]. In addition, the vernier effect can be realized with higher-order MRRs of varying radii, which allows for exponential increases in FSR [39]. Similarly, adiabatic MRRs with larger FSRs [14, 40, 41], or even FSR-free MRRs [42, 43] could serve as effective switching elements. An increase in FSR would enable the SWSS to accommodate more wavelength channels while maintaining low crosstalk, thereby enhancing both its scalability and overall performance.

**Table 3.** Performance comparison with other reported integrated switches

| Reference | Scale | Elements Number | Platform | On-chip Loss | The worst Crosstalk |
|---|---|---|---|---|---|
| [11] | 4×4×1λ | 32 | Si/SiN | 1.8-20.4 dB | -50 dB |
| [14] | 1×4×4λ | 16 | SOI | <1 dB | -20 dB |
| [18] | 8×8×8λ | 128 SOA gates +8 AWGs | InGaAsP/InP | 7.6-14.7 dB | Penalty: 3.3 dB |
| [20] | 8×4×2λ | 64 | SOI | 6-14 dB | -20/-32 dB |
| [33] | 2×2×2λ | 4 | SOI | Avg. 5.1 dB | -21 dB |
| This work | 4×4×4λ | 32 | SOI | 2.3-8.6 dB | -35.3 dB |

## 5. Conclusion

In this paper, we propose a new type of dilated crosspoint topology for space-and-wavelength selective switching and demonstrate an ultra-compact 4×4×4λ SWSS. This innovative design significantly reduces the number of required switching elements, achieving a substantial reduction compared to existing SWSS designs. The proposed SWSS benefits from a compact design with only a 1 × 2.25 mm$^2$ footprint and efficient use of switching elements with only 32 MRRs. This SWSS achieves efficient and flexible routing of optical signals across four wavelength channels. Experimental results reveal that the switch exhibits a low insertion loss ranging from 2.3 dB to 8.6 dB and crosstalk levels between -35.3 dB and -59.7 dB, with the 3dB bandwidth of over 20 GHz for all four wavelength channels. Overall, the demonstrated SWSS offers a powerful yet highly-scalable solution for data-intensive applications in data centers.

## Appendix A

An $N \times N \times N\lambda$ dilated crosspoint SWSS with $N$ columns can have blocking states when operating under fully-loaded conditions. We simplify this problem mathematically by populating the inputs and outputs of a SWSS in a square and denoting its wavelength channel with numbers. An example of a blocking instance is shown in Fig. A1 (a). The square corresponds to the schematic of the crosspoint structure, as shown in Fig. 6(a). In the first-stage crossbar, the first column corresponds to inputs $I_1$ to $I_4$, while the last column of the second-stage crossbar corresponds to outputs $O_1$ to $O_4$. The numbered elements (1–4) in the square represent the MRRs positioned accordingly, operating at resonant channels $C_1$ to $C_4$. In this instance, signals in $C_1$ are routed from $I_1$ to $O_2$ while $C_2$, $C_3$ and $C_4$ are routed to $O_1$. Clearly, the element marked with "B" can not be tuned to any wavelength channel without disrupting the routing paths within the same channel. Therefore, to ensure non-blocking operations of the SWSS under full load, each stage of the crossbar must be represented as a Latin Square. The global switching states equal to the square of the Latin Square count, which can be obtained from a look-up table [25]. To exclude redundant switching states, several scenarios need to be considered. The first is illustrated in Fig. A1 (b). An $l \times l$ sub-square is selected where $l \leq N/2$. When routing signals from the input port to the corresponding output port or inverse output port through this sub-square, the permutation remains unchanged if we simultaneously change the column order of the corresponding rows in both the first and second stage crossbars. Consequently, such a state is considered redundant. The number of these situations can be expressed with the following equation:

$$S_{r1}(N) = \prod_{l \in H}\left(C_N^l \cdot l!\right). \tag{A1}$$

Here, set $H$ consists of all the positive integer smaller than $N/2$. As we need to consider another situation if $l$ is the the divisor of $N$, so divisors should also be excluded.

In the second scenario, an $i \times i$ sub-square is selected from the Latin Square. An instance of $N=8$ is illustrated in Fig. A1(c). When $i=2$, following the same process as in the first scenario, a new square is formed, marked with red lines. These 2×2 squares act as the elements of larger 2×2 squares, which also exhibit similar redundant situations. The redundant states in this case can be represented by:

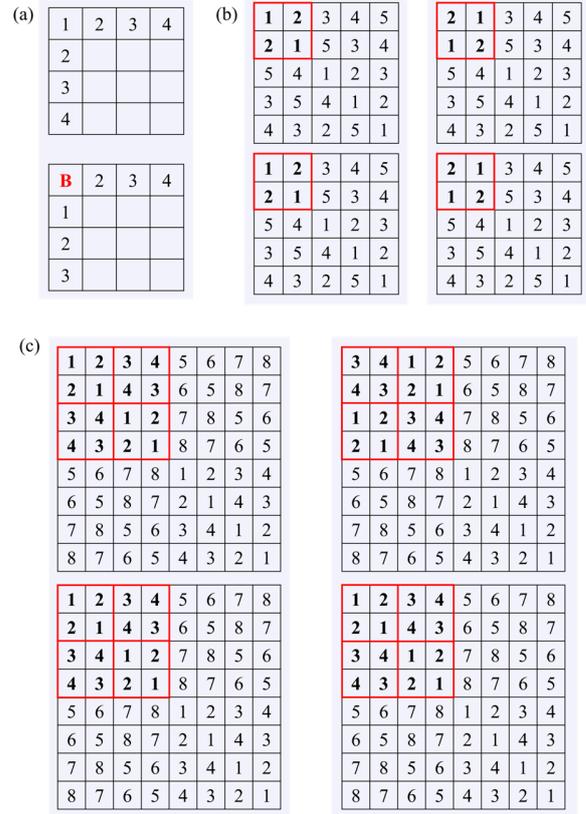

**Fig. A1.** Diagrams of the switching states.

$$S_r(N) = \prod_{i \in D1}\left(\left(C_N^i \cdot i!\right) \cdot \prod_{j \in D2}\left(C_{\frac{N}{i}}^j \cdot j!\right)\right). \tag{A2}$$

In this equation, the set $D1$ consists of all divisors of $N$, and the set $D2$ consists of divisors of $N/i$. Both $D1$ and $D2$ exclude 1. Note that $N$ and $N/i$ should also be included in these sets. Thus, considering all redundant switching states, the total number of redundant states is derived as:

$$S_r(N) = \prod_{i \in D1}\left(\left(C_N^i \cdot i!\right) \cdot \prod_{j \in D2}\left(C_{\frac{N}{i}}^j \cdot j!\right)\right) \cdot \prod_{l \in H}\left(C_N^l \cdot l!\right). \tag{A3}$$

The comparison between the effective switching states and the number of permutations was discussed in Section 2, where it was shown that reducing the number of wavelength channels can achieve wide-sense non-blocking conditions. However, after reducing the wavelength channels, a correction term must be applied. If the SWSS reduces $W$ wavelength channels, the original elements corresponding to $N$, $N$-1 … $N$-$W$ become redundant and should no longer be treated as distinct elements. Consequently, a correction factor of $W!$ must be incorporated. Therefore, the wide-sense non-blocking evaluation criterion should compare $P$ with $S_e/W!$.

**Funding.** This work was funded by the UK Research and Innovation, Engineering and Physical Sciences Research Council (UKRI-EPSRC), project QUDOS (EP/T028475/1); and the European Union's Horizon Europe Research and Innovation Program, project PUNCH (101070560); and the European Union's

Horizon Europe Research and Innovation Program, project INSPIRE (101017088).

**Disclosures.** The authors declare no conflicts of interest.
## References

[1] A. Singh, J. Ong, A. Agarwal, *et al.*, Jupiter rising: A decade of clos topologies and centralized control in google's datacenter network," ACM SIGCOMM Comp. Com., **45**(4), 183-197 (2015).

[2] Q. Cheng, M. Bahadori, M. Glick, *et al.*, "Recent advances in optical technologies for data centers: a review," Optica, **5**(11), no. 11, 1354-1370 (2018).

[3] X. Chen, J. Lin, and K. Wang, "A Review of Silicon‐Based Integrated Optical Switches," Laser & Photonics Rev., **17**(4), 2200571 (2023).

[4] M. Stepanovsky, "A comparative review of MEMS-based optical cross-connects for all-optical networks from the past to the present day," IEEE Commun. Surv. Tutor., **21**(3), 2928-2946 (2019).

[5] W. M. Mellette, G. M. Schuster, G. Porter, *et al.*, "A scalable, partially configurable optical switch for data center networks," J. Light. Technol., **35**(2), 136-144 (2016).

[6] W. M. Mellette and J. E. Ford, "Scaling limits of MEMS beam-steering switches for data center networks," J. Light. Technol., **33**(15), 3308-3318 (2015).

[7] A. Komar, R. Paniagua-Domínguez, A. Miroshnichenko, *et al.*, "Dynamic beam switching by liquid crystal tunable dielectric metasurfaces," ACS Photonics, **5**(5), 1742-1748 (2018).

[8] M. S. Li, A. Y. Fuh, and S. Wu, "Optical switch of diffractive light from a BCT photonic crystal based on HPDLC doped with azo component," Opt. Lett., **36**(19), 3864-3866 (2011).

[9] R. Soref, "Tutorial: Integrated-photonic switching structures," APL Photonics, **3**(2), 021101 (2018).

[10] Q. Cheng, C. Yao, N. Calabretta, et al. "Photonic switch fabrics in data center/high-performance computing networks." Integrated Photonics for Data Communication Applications. Elsevier,. 265-301 (2023).

[11] Q. Cheng, L. Y. Dai, N. C. Abrams, *et al.*, "Ultralow-crosstalk, strictly non-blocking microring-based optical switch," Photon. Res., **7**(2), 155-161 (2019).

[12] W. Gao, X. Li, L. Lu, *et al.*, "Broadband 32×32 Strictly‐Nonblocking Optical Switch on a Multi‐Layer $Si_3N_4$‐on‐SOI Platform," Laser & Photonics Rev., **17**(11), 2300275 (2023).

[13] C. Vagionas, A. Tsakyridis; T. Chrysostomidis, *et al.*, "Lossless 1× 4 Silicon Photonic ROADM Based on a Monolithic Integrated Erbium Doped Waveguide Amplifier on a $Si_3N_4$ Platform," J. Light. Technol., **40**(6), 1718-1725 (2022).

[14] C. Zhang, Y. Xiang, S. Liu, *et al.*, "Silicon Photonic Wavelength-Selective Switch Based on an Array of Adiabatic Elliptical-Microrings," J. Light. Technol., **41**(17), 5660-5667 (2023).

[15] Y. Ma, L. Stewart, J. Armstrong, *et al.*, "Recent Progress of Wavelength Selective Switch," J. Light. Technol., **39**(4), 896-903 (2021).

[16] S. Araki, Y. Suemura, N. Henmi, *et al.*, "Highly scalable optoelectronic packet-switching fabric based on wavelength-division and space-division optical switch architecture for use in the photonic core node [Invited]," J. Opt. Netw., **2**(7), 213-228 (2003).

[17] R. P. Luijten and R. Grzybowski, "The OSMOSIS Optical Packet Switch for Supercomputers," in *Optical Fiber Communication Conference (OFC)* (2009), paper OTuF3,

[18] R. Stabile, A. Rohit, and K. A. Williams, "Monolithically Integrated 8 × 8 Space and Wavelength Selective Cross-Connect," J. Light. Technol., **32**(2), 201-207 (2014).

[19] Q. Cheng, M. Bahadori, M. Glick, *et al.*, "Scalable Space-and-Wavelength Selective Switch Architecture Using Microring Resonators," in *Conference on Lasers and Electro-Optics (CLEO)* (2019), paper STh1N.4.

[20] A. S. P. Khope, M. Saeidi, R. Yu, *et al.*, "Multi-wavelength selective crossbar switch," Opt. Express, **27**(4), 5203-5216 (2019).

[21] L. Luo, R. Ma, R. V. Penty, *et al.*, "Unlocking Electro-optic Resonant Phase Shifting for Multi-dimensional, Ultra-dynamic Photonic Switches," arXiv preprint, arXiv:2403.02866 (2024).

[22] A. W. Poon, X. Luo, F. Xu, *et al.*, "Cascaded Microresonator-Based Matrix Switch for Silicon On-Chip Optical Interconnection," Proc. IEEE, **97**(7), 1216-1238 (2009).

[23] Q. Cheng, Y. Huang, H. Yang, *et al.*, "Silicon Photonic Switch Topologies and Routing Strategies for Disaggregated Data Centers," IEEE J. Quantum Electron., **26**(2), 1-10 (2020).

[24] P. DasMahapatra, R. Stabile, A. Rohit, *et al.*, "Optical Crosspoint Matrix Using Broadband Resonant Switches," IEEE J. Quantum Electron., **20**(4), 1-10 (2014).

[25] D. S. Stones, "The many formulae for the number of Latin rectangles," Electron. J. Comb., **17**, A1-A1 (2010).

[26] A. S. P. Khope, S. Liu, Z. Zhang, *et al.*, "2 lambda switch," Opt. Lett., **45**(19), 5340-5343 (2020).

[27] K. Ikeda, R. Konoike, and K. Suzuki, "Large-Scale and Multiband Silicon Photonics Wavelength Cross-Connect Switch With FSR-Free Grating-Assisted Contra-Directional Couplers," J. Light. Technol., **42**(12), 4310 - 4316 (2024).

[28] W. Bogaerts, P. D. Heyn, T. V. Vaerenbergh, *et al.*, "Silicon microring resonators," Laser & Photonics Rev., **6**(1), 47-73 (2012).

[29] M. Bahadori, M. Nikdast, S. Rumley, *et al.*, "Design Space Exploration of Microring Resonators in Silicon Photonic Interconnects: Impact of the Ring Curvature," J. Light. Technol., **36**(13), 2767-2782 (2018).

[30] Z. Li, L. Zhou, Liangjun Lu, S. Zhao, *et al.*, "4 × 4 nonblocking optical switch fabric based on cascaded multimode interferometers," Photon. Res. **4**(1), 21-26 (2016)

[31] J. Song, Q. Fang, S. H. Tao, *et al.*, "Fast and low power Michelson interferometer thermo-optical switch on SOI," Opt. Express **16**, 15304-15311 (2008)

[32] N. Ning, Q. Zhang, Q. Huang, *et al.*, "Thermal flux manipulation on the silicon photonic chip to suppress the thermal crosstalk" APL Photonics, **9**(4), 046108 (2024).

[33] Y. Huang, Q. Cheng, A. Rizzo, *et al.*, "Push-pull microring-assisted space-and-wavelength selective switch," Opt. Lett., **45**(10), 2696-2699 (2020).

[34] Z. Wan, Q. Cen, Y. Ding, *et al.*, "Virtual-State Model for Analyzing Electro-Optical Modulation in Ring Resonators," Phys. Rev. Lett., **132**(12), 123802 (2024).

[35] H. Cai, S. Fu, Y. Yu, *et al.*, "Lateral-zigzag pn junction enabled high-efficiency silicon micro-ring modulator working at 100gb/s," IEEE Photonics Technol. Lett., **34**(10), 525-528 (2022).

[36] P. Chen, S. Chen, X. Guan, *et al*, "High-order microring resonators with bent couplers for a box-like filter response," Opt. Lett., **39**(21), 6304-7 (2014).

[37] P. DasMahapatra, R. Stabile, A. Rohit, *et al.*, "Optical Crosspoint Matrix Using Broadband Resonant Switches," IEEE J. Sel. Top. Quantum Electron., **20**(4), 1-10 (2014).

[38] Y. Huang, Q. Cheng; Y. Hunget, *et al.*, "Multi-Stage 8 × 8 Silicon Photonic Switch Based on Dual-Microring Switching Elements," J. Light. Technol., **38**(2), 194-201 (2020).

[39] R. Boeck, N. A. Jaeger, N. Rouger, *et al.*, "Series-coupled silicon racetrack resonators and the Vernier effect: theory and measurement," Opt. Express, **18**(24), 25151-25157 (2010).

[40] G. Liang, H. Huang, A. Mohanty, *et al.*, "Robust, efficient, micrometre-scale phase modulators at visible wavelengths," Nat. Photon., **15**(12), 908-913 (2021).